\documentstyle[preprint,aps,prl]{revtex}

\begin{document}
\title
{\Large{\bf Wavelet Based Fractal Analysis of Airborne Pollen}}

\author{\bf M.E. Degaudenzi, C.M. Arizmendi}

\address{Depto. de F\'{\i}sica, Facultad de
Ingenier\'{\i}a, Universidad Nacional de  Mar del
Plata,\\ Av. J.B. Justo 4302, 7600 Mar del Plata,
Argentina\\}

\vspace {1 truecm}

\date{\today}

\maketitle

\begin{abstract} 
\mediumtext

The most abundant biological particles in the atmosphere are pollen grains
and spores.
Self protection of pollen allergy is possible through the information
of future pollen contents in the air. In spite of the importance of airborne pollen concentration forecasting,
it has not
been possible to predict the pollen concentrations with great accuracy, and
about 25\% of the daily pollen forecasts have resulted in failures.
 Previous analysis of the dynamic characteristics of
atmospheric pollen time series indicate that the system can be described by
a low dimensional chaotic map.
We apply the wavelet transform to study the multifractal characteristics of an airborne pollen time series.
We find the persistence behaviour associated to
low pollen concentration values and to the most rare events of highest pollen concentration values. The information and the correlation dimensions correspond to a chaotic system showing loss of information with time evolution.

\end{abstract}
\newpage

\section{Introduction}
Pollen allergy is a common disease causing hay fever in 5-10\%
of the population. Although not a life threatening disease,
the symptoms can be very troublesome, furthemore, the costs to the
social sector due to pollen related diseases are high.
Self protection of hay fever patients is possible through the information
of future pollen contents in the air \cite{leuch}.

Models to forecast pollen concentration in the air are principally based on
pollen and atmospheric weather interactions. Several statistical techniques
\cite{comt,orro,gold}, have been used to predict future
atmospheric pollen concentrations from weather conditions of
the day and of recent previous days. In spite of these attempts, it has not
been possible to predict the pollen concentrations with great accuracy, and
about 25\% of the daily pollen forecasts have resulted in failures
\cite{gold}. \\
A reason of these failures could be that the methods used in airborne
pollen forecasting
are based in standard linear statistical techniques which don't
suit when the phenomenon to forecast is esentially non-linear.

A previous analysis of the dynamic characteristics of time series of 
atmospheric pollen was developed by Bianchi et al. \cite{bian}, through
the study
of the correlation dimension \cite{gras,abra}. The dimension
found was a low and non integer value \cite{bian},
which indicates that the system may be described by a nonlinear
function of just a few variables relating nearest pollen
concentrations of the time series. The fact that the correlation dimension
found
was fractal predicts that this function, also called map in nonlinear
dynamics
can display chaotic behavior under certain circumstances.
The existence of a low dimensional map suggests possibilities for
short-term
prediction
\cite {farm} through the use of some nonlinear model.
Artificial neural networks have been widely used to predict future values
of chaotic time series identifying the nonlinear model by extracting
knowledge from the past \cite{lape}.
Very good pollen concentrations forecasts were obtained  using 
neural nets \cite{ariz} and, in a previous work, the
hypothesis that random fluctuations appearing in the pollen time series
are
produced by Gaussian noise was rejected \cite{ariz1}. 

To continue with the characterization of airborne pollen concentrations
the next step would be to study what kind of correlation is associated
with its fluctuations.
The Hurst exponent $H$ is broadly used as a measure of the persistence of
a statistical phenomenon.
$0\leq H <0.5$ points an antipersistent time series, commonly driven by a
phenomenon called ``Noah Effect" (If you see the Bible, the storm changed
everything in a moment).  It characterizes a system that reverses itself
more frequently and covers less distance than a random walk. 
$0.5<H\le1 $ implies that we are analyzing a persistent time series which
obeys to the ``Joseph Effect" (In the Bible refers to 7 years of loom,
happiness and health and 7 years of hungry and illness).  This system has
long memory effects: what happens now will influence the future, so there
is a very deep dependence with the initial conditions. Persistent
processes are common in nature \cite{hein,pete}.
 If the distribution is homogeneous there is an unique $\alpha=H$, but if
it is not there are several exponents $\alpha$.  The most frequent
$\alpha$ will characterize the series and will play as Hurst exponent $H$.
A very efficient new method to obtain the $f(\alpha)$ singularity spectrum
of a pollen time series relies on the use of a mathematical tool
introduced in signal analysis in the early eighties: the {\sl wavelet
transform}. The wavelet transform has been proved very efficient to detect
singularities and fractals are singular functions indeed. Arneodo {\sl et
al}\cite{arne,stru} developed the {\sl wavelet transform modulus
maxima} (WTMM)
method as a technique to study fractal objects. In this method the
wavelet is used as an oscillating variant of the $``$square'' function of
a box. WTMM was succesfully applied to study fractal properties of diverse
systems such as DNA nucleotide sequences \cite{dna,dna1}, Modane
turbulent velocity signal \cite{arne,turb} and a cool
flame experiment \cite{cool}. 
We apply WTMM to obtain the Hurst exponents $H$ associated with the pollen time series as a whole as well as the persistence of the important rare peaks of highest concentrations. Another important tool in describing multifractals that are obtained through WTMM are the generalized fractal dimensions $D_q$.

\section{Experimental Setup}

The material used in this work was from our chaos study of pollen
series \cite{bian}.

Data of airborne pollen concentration were  obtained   with   an
automatic and volumetric Burkard pollen and spore   trap,   situated
at the roof of the Facultad de Ciencias Exactas y Naturales of our
University, 12
meters above ground level. The   area   surrounding   the   sample   is
typical of Mar del Plata.
The great distance from the sampling site to the emission sources makes
the particular emission spectra not important.

          Ten liters of air per minute were sucked through a 14 x 2
$mm^{2}$
orifice, always orientated against the wind flow. The sucking rate
is checked weekly. Behind the slit, a drum rotates at a speed of 2
$mm$ per hour. The particles are   collected   on   a   cellophane   tape
(Melinex),   19  $ mm$   wide,   just   below   the   orifice.   The  
sticky
collecting surface comprises nine parts vaseline: one   part   paraffin  
in
toluene. The exposed tape is   removed   from   the   drum,   cut   into
pieces of 48 mm, corresponding   to   24-h   intervals, then
embedded into a solution of polivinylalcohol (Gelvatol), water   and
glycerol and covered with a cover glass. Slides were   studied
as 12 transects per day. The pollen was counted at a magnification of X400
for   the first year cycle (August 1987-8)   and at X200 for the   second
(August 1988-9), and corresponding to 13.5 and 27 min of sampling
every 2 h respectively.   The   method   of   counting   pollen
follows that of K\H{a}pyl\H{a}  and   Penttinen \cite{pent}.
 Hourly   counts   were
stored in a database file   for   further   analysis.   Statistics   of
hourly counts may be seen in Table 1 and 2 of \cite{bian}.
The concentration values correspond to total pollen grains.  The main
species found were: Cupressus, Gramineae, Eucalyptus, Pinace,
Chenopodiineae, Plantago, Cyperaceae, Betula, Cruciferae, Compositae
Tueulflorae, Ambrosia, Ulmus, Umbelliferae, Platanus and Fraxinus.

\section{The multifractal formalism}
The aim of this formalism is to determinate the f($\alpha$) singularity
spectrum of a measure $\mu$ .  It associates the Haussdorff Dimension of
each point with the singularity exponent $\alpha$, which
gives us an idea of the strength of the singularity.
 
\begin{equation}
N_{\alpha}(\epsilon) \sim \epsilon^{-f(\alpha)}, 
\end{equation}

where $N_{\epsilon}$ is the number of boxes needed to cover the measure
and
$\epsilon$ is the size of each box \cite{hein}.

A partition function $Z$ can be defined from this spectrum (it is the same
model as the thermodinamic one).

\begin{equation}
Z(q,\epsilon) = \sum_{i=1}^{N(\epsilon)}\mu_i^q(\epsilon) \sim
\epsilon^{\tau(q)}  \:\:\:\:for \: \epsilon  \to 0,
\end{equation}

where $\tau(q)$ is a spectrum which arouses by Legendre transforming the
$f(\alpha)$ singularity spectrum.

The spectrum of $generalized\: fractal\: dimensions\: D_q$ is obtained
from
the spectrum $\tau(q)$

\begin{equation}
D_q = \frac{\tau(q)}{(q-1)},
\end{equation}

The capacity or box dimension of the support of the distribution is given
by $D_0=f(\alpha(0))= -\tau(0)$.
$D_1=f(\alpha(1))=\alpha(1)$ corresponds to the scaling behavior of the
information and is called $information\: dimension$. 
For $q\ge2$, $D_q$ and the $q-point\: correlation\: integrals$ are
related. 

As we will show in the following section the Wavelet Transform is specially
suited to analyze a time series as a multifractal. 
 
\section{Wavelet Transform}

The Wavelet Transform (WT) \cite{goup,gros} of a signal $s(t)$ consists
in decomposing it into
frequency and time coefficients, asociated to the wavelets.  The analyzing
wavelet $\psi$, by means of translations and dilations, generates the so
called family of wavelets.

The Wavelet Transform turns the signal $s(t)$ into a function
$T_\psi[s](a,b)$:

\begin{equation}
T_\psi[s](a,b)= \frac{1}{a}\int {\psi}^{*}[\frac{t-b}{a}]s(t)dt,
\end{equation}
where ${\psi}^{*}$ is the complex conjugate of $\psi$, $a$ is the
frequency
dilation factor and $b$, the time translation parameter.

The wavelet to apply must be chosen with the condition:
\begin{equation}
\int  \psi(t)dt=0,
\end{equation}

and to be orthogonal to lower order polynoms
\begin{equation}
\int  t^m\psi(t)dt=0 \:\:\:\:\:\:\: 0\le m\le n;
\end{equation}  
where $m$ is the order of the polynom.  In other words, lower order
polinomial behavior is eliminated and we can detect and characterize
singularities even if they are masked by a smooth behavior.

The WT provides a useful tool to the detection of self-similarity or
self-affinity in temporal series. For a value $b$ in the domain of the
signal, the modulus of the transform is maximized when the frequency $a$
is
of the same order of the characteristic frequency of the signal $s(t)$ in
the neighborhood of $b$, this last one will have a local singularity
exponent  $\alpha(b)\: \in \:]n,n+1[$. 
 
This means that around $b$ 

\begin{equation}
|s(t)-P_n(t)| \sim |t-b|^{\alpha(b)},
\end{equation}

where $P_n(t)$ is an $n$ order polynomial, and

\begin{equation}
T_{\psi}(a,b) \sim a^{\alpha(b)},
\end{equation}

provided the first $n+1$ moments are zero.  

If we have $\psi^{(N)}= d^{(N)}(e^{x^2/2})/dx^{N}$, the first $N$ moments
are vanishing. 

The Wavelet Modulus Function $|T_\psi[s](a,t)|$ will have a local maximum
around the points where the signal is singular.
These local maxima points make a geometric place called modulus maxima
line ${\cal L}$.

\begin{equation}
|T_\psi[s](a,b_l(a))| \sim a^\alpha(b_l(a))\:\:\:\:\: for \:\: a \to 0,
\end{equation}

where $b_l(a)$ is the position at the scale $a$ of the maximum belonging
to
the the line ${\cal L}$.

The Wavelet Transform Modulus Maxima Method (WTMM) consists in the analysis of
the
scaling behavior of some partition functions $Z(q,a)$ that can be defined
as:

\begin{equation}
Z(q,a) = {\sum}|T_\psi[s](a,b_l(a))|^q,   
\end{equation}

and will scale like $a^{\tau(q)}$ \cite{arne,stru}.

This partition function works like the previously defined partition function
for singular measures.  For $q>0$ will prevail the most pronounced modulus
maxima and, on the other hand, for $q<0$ will survive the lower ones.
The most pronounced modulus take place when very deep singularities are
detected, while the others correspond to smoother singularities.
We can get $\tau(q)$ (Eq. 2) and 
obtain $f(\alpha)$ and $D_q$ spectra, as explained previously.
The shape of $f(\alpha)$ is a hump that has a maximum value, $\alpha$
corresponding to this maximum may be associated with the general behavior
of the series. So, this
particular singularity exponent can be thought like the Hurst exponent $H$
for the series as a whole.  

\section{Application of WTMM to the pollen time series}
The airborne pollen concentration time series may be seen in Fig. 1.  The
third derivative of Gaussian function was chosen as analyzing wavelet:

\begin{equation}
\psi^{(3)}(t)= {d^3 \over dt^3} (e^{t^2/2}),
\end{equation}

Twelve wavelet transform data files were obtained applying the Wavelet
Transform with $\psi^{(3)}$, ranging the scaling factor $a$ from
$a_{min}=1/256$ to $a_{max}=8$ in steps of $2^n$. To give an idea of the effect of the change of scale on wavelet transform of the pollen time series, three of them are shown in Fig. 2. 

Then we computed the partition function $Z(q,a)$ for $-30\le q\le30$ and $1/256\le a \le 8$,
getting $\tau(q)$, as shown in Fig. 3.

$\tau(q)$ is a nonlinear convex increasing function with $\tau(0)=-0.97$
and two asymptotic slopes which are $\alpha_{min}=0.40$ for $q>0$ and
$\alpha_{max}=1.39$ for $q<0$.

This lays the corresponding $f(\alpha)$ singularity spectrum obtained by
Legendre transforming $\tau(q)$ that is displayed in Fig. 4. The single humped shape with a nonunique H\"older exponent obtained characterizes a multifractal. 

As expected from $\tau(q)$, the support of $f(\alpha)$ extends over a
finite interval which bounds are $\alpha_{min}=0.40$ and
$\alpha_{max}=1.39$. 
The minimum value, $\alpha_{min}$, corresponds to the strongest
singularity
which characterizes the most rarified zone, whereas higher values exhibit
weaker
singularities until $\alpha_{max}$ or weakest singularity which
corresponds
to the densiest zone.  $\alpha_{min}<1/2$ corresponds to an
antipersistent process and $\alpha_{max}>1$, to a regular process.

The $D_q$ spectrum obtained from $\tau(q)$ can be seen in Fig. 5.
The support dimension $D_o=D_{max}=-\tau(0)=0.97$; which implies that the
capacity of the support is approximately 1; i.e. the support is not a
fractal. $D_q$ converges asymptotically to $\alpha_{min}=0.40$ for
$q_{max}$ and to $\alpha_{max}=1.38$ for $q_{min}$.

The H\"older exponent for the dimension support, $\alpha(D_{max})$, is
0.90. This particular $\alpha$ corresponds to $f(\alpha)_{max}$ or
$D_{max}$ which implies that the sucesses with
$\alpha=\alpha(D_{max})=0.90$ are the most frequent ones.  
$0.5<\alpha\le1 $ implies we are analyzing a persistent time series which
obeys to the ``Joseph Effect'' (In the Bible refers to 7 years of loom,
happiness and health and 7 years of hungry and illness) \cite{pete}.  This
system has long memory effects: what happens now will influence the
future,
so there is a very deep dependence with the initial conditions.  It may be
thought like a Fractional Brownian Motion of $\alpha>0.5$. 
A Hurst exponent of 0.90 describes a very persistent time series, what is
expected in a natural process involved in an inertial system. 
$\alpha$ can be known as H\"older Exponent or Singularity Exponent, too.
If the distribution is homogeneous there is an unique $\alpha=H$ (for
example Fractional Brownian Motion), but if it is not there are several
exponents $\alpha$.  The most frequent $\alpha$ will characterize the
series and will play as Hurst exponent.

$\stackrel{-}{\alpha}=(\alpha_{min}+\alpha_{max})/2=\alpha(D_{max})=0.90$.
This means that the curve is equally humped in both sides with the
consequence of having the same inhomogeneity in the less frequent events
associated with the $q<0$ branch and in the more frequent ones associated
with the $q\ge0$ branch.
 
The information dimension is $D_1=f(\alpha(1))=f(0.68)=0.68$ which
features
the scaling behavior of the information.  It plays an important role in
the
analysis of nonlinear dynamic systems, specially in describing the loss of
information as chaotic system evolves in time \cite {manf}.  $D_1=0.68$
implies that we
are in the presence of a chaotic system. 

The correlation dimension is $D_2=0.60$ which characterizes a chaotic
attractor and is very close to the value previously obtained with the
Grassberger-Procaccia method \cite{bian}.

\section{Conclusion}

The Wavelet Transform Modulus Maxima Method was applied to study the multifractal characteristics of an airborne pollen time series.
We have found that pollen time series behave as a whole like long term memory persistent
phenomena , as most ones in nature.  The most common events associated with $\alpha_{max}$ which correspond to low pollen concentration values behave in a persistent way as the whole series. On the other hand, the most rare events associated in the multifractal formalism to $\alpha_{min}$ which correspond to highest pollen concentration values behave in an antipersistent way characterized by the ``Noah Effect'', changing suddenly and catastrophically the air conditions. Both the information and the correlation dimensions correspond to a chaotic system showing loss of information with time evolution. 

\section{Acknowledgments}
 C.M.A. would like to thank Alain Arneodo for
introducing him to wavelet transform multifractal analysis . This work
was partially supported by a grant from the Universidad Nacional de Mar del Plata.

\newpage
FIGURE CAPTIONS

\vspace{0.4cm}
Fig. 1.Two years of airborne pollen concentration time series. The time step is 2 hours.

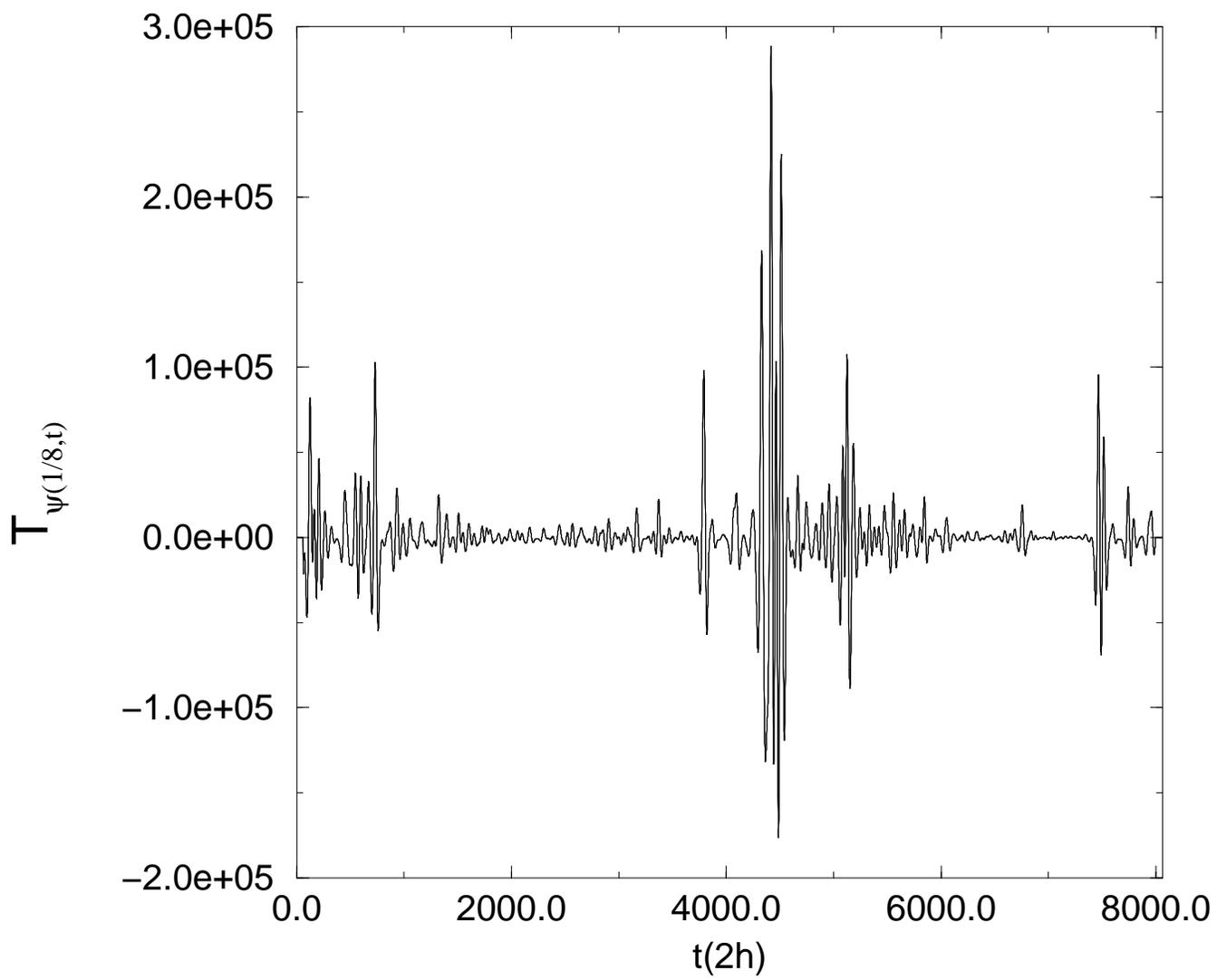
Fig. 2. Wavelet transform data of pollen time series distribution. a) scale $a=1$, b)  scale $a=1/8$, c) scale $a=1/64$.  

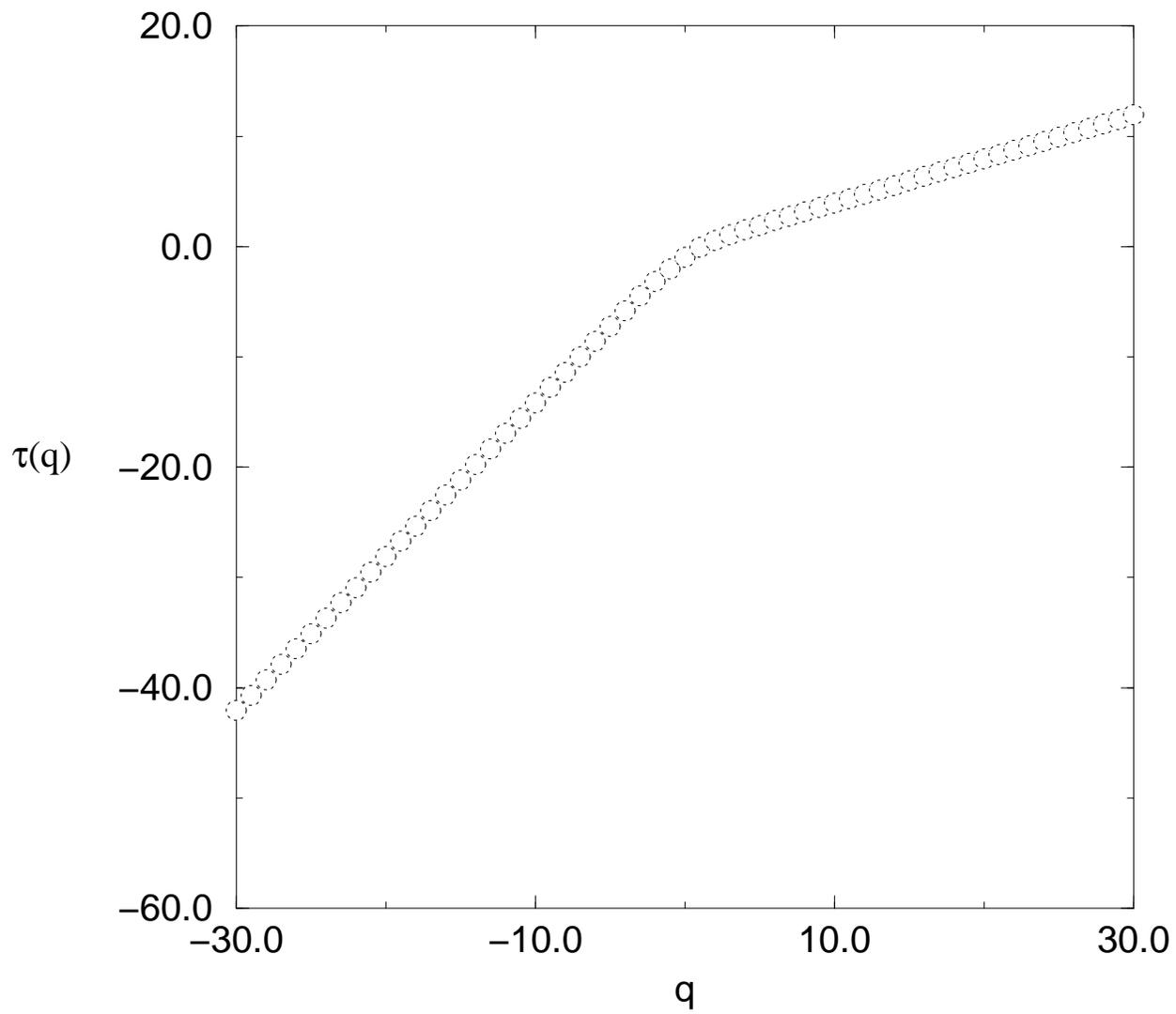
Fig. 3. $\tau(q)$ spectrum of pollen time series.

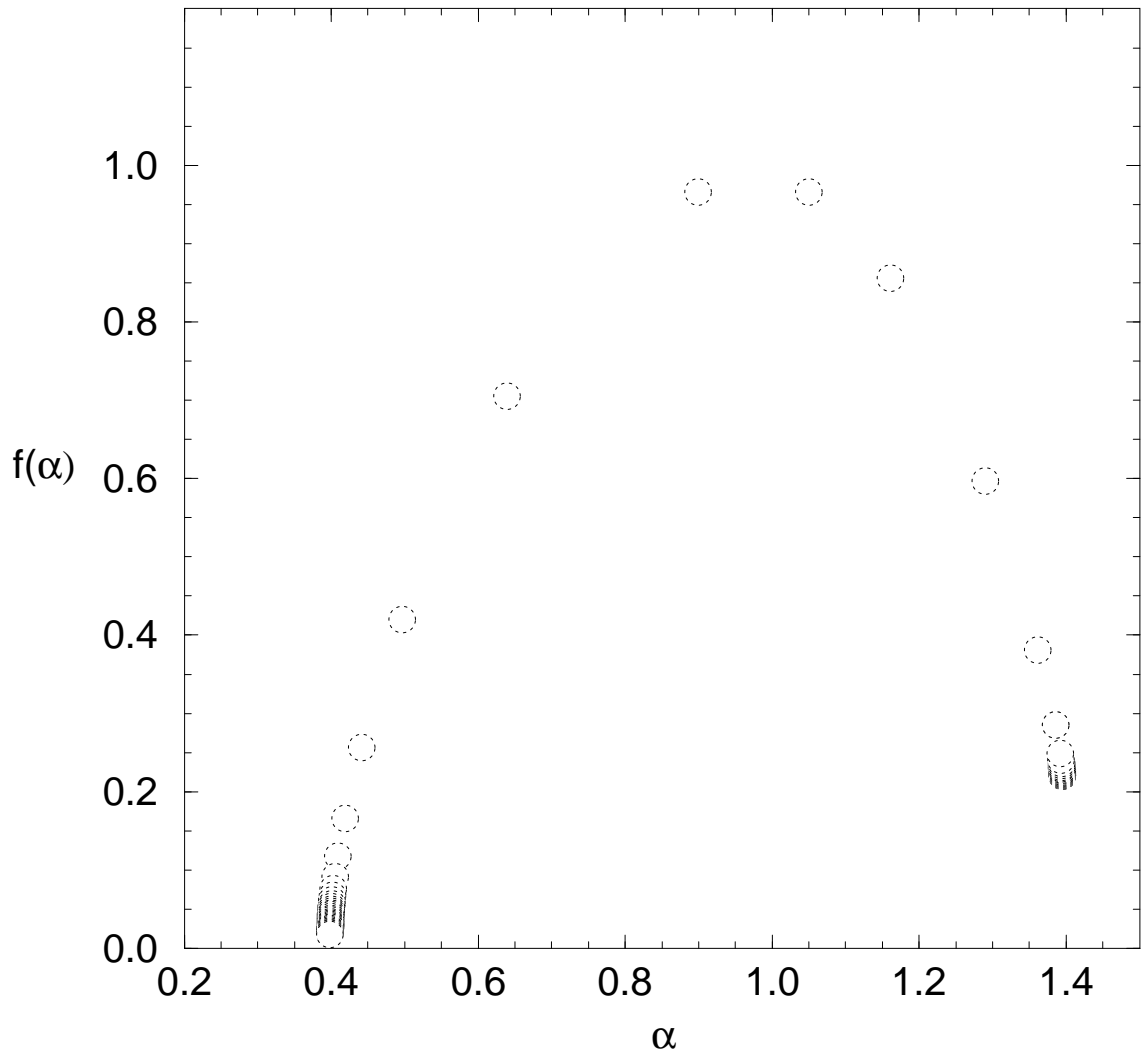
Fig. 4. $f(\alpha)$ spectrum of pollen time series. the support of $f(\alpha)$ bounds are $\alpha_{min}=0.40$ which corresponds to an
antipersistent process and
$\alpha_{max}=1.39$ to a regular process. The H\"older exponent for the dimension support, $\alpha(D_{max})$, is
0.90 which characterizes a persistent process.

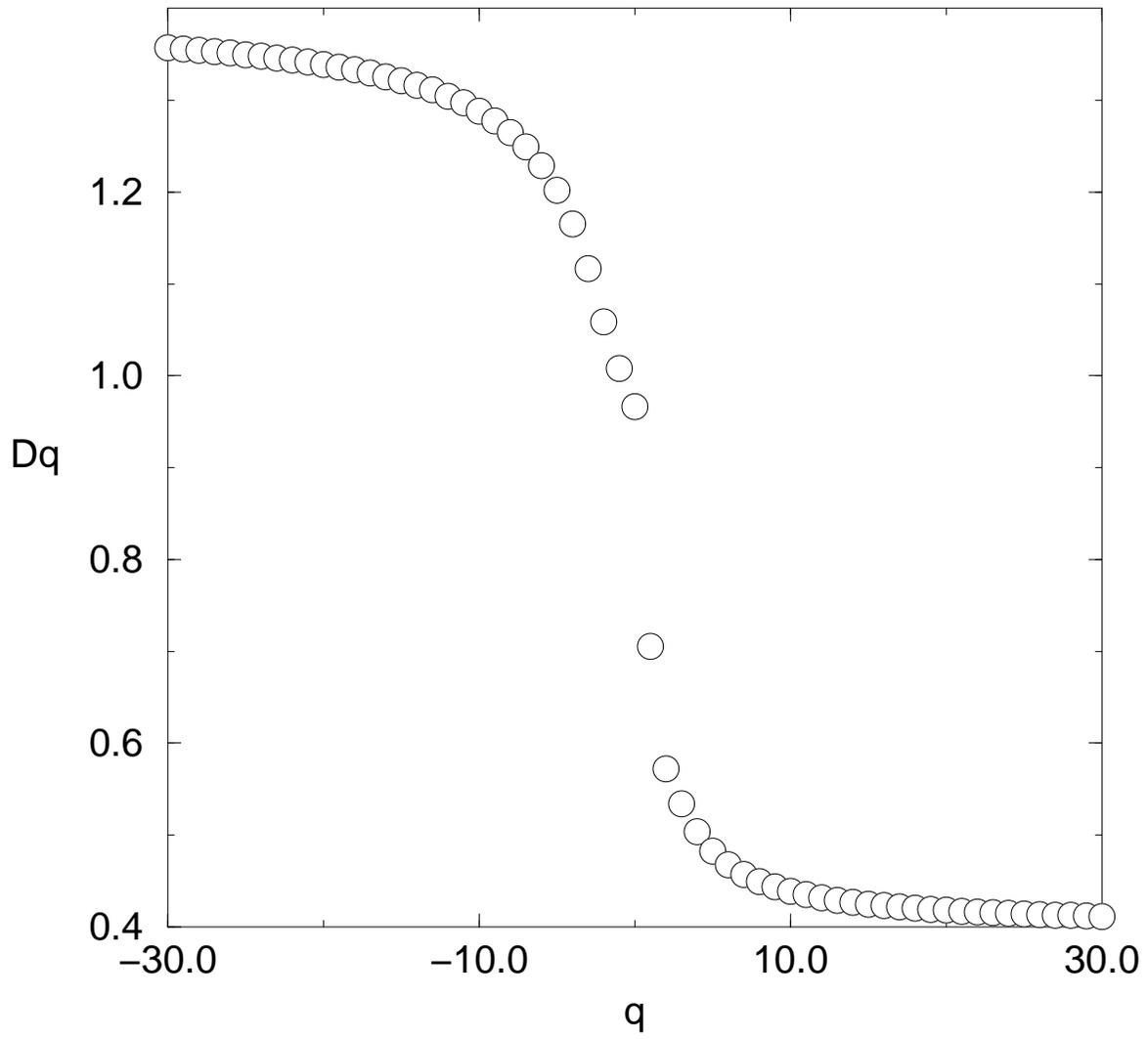
Fig. 5. $D_q$ spectrum of pollen time series. The support dimension $D_o=D(q=0) \sim 1$.  $D_q$ converges asymptotically to $\alpha_{min}=0.40$ for
$q_{max}$ and to $\alpha_{max}=1.38$ for $q_{min}$.

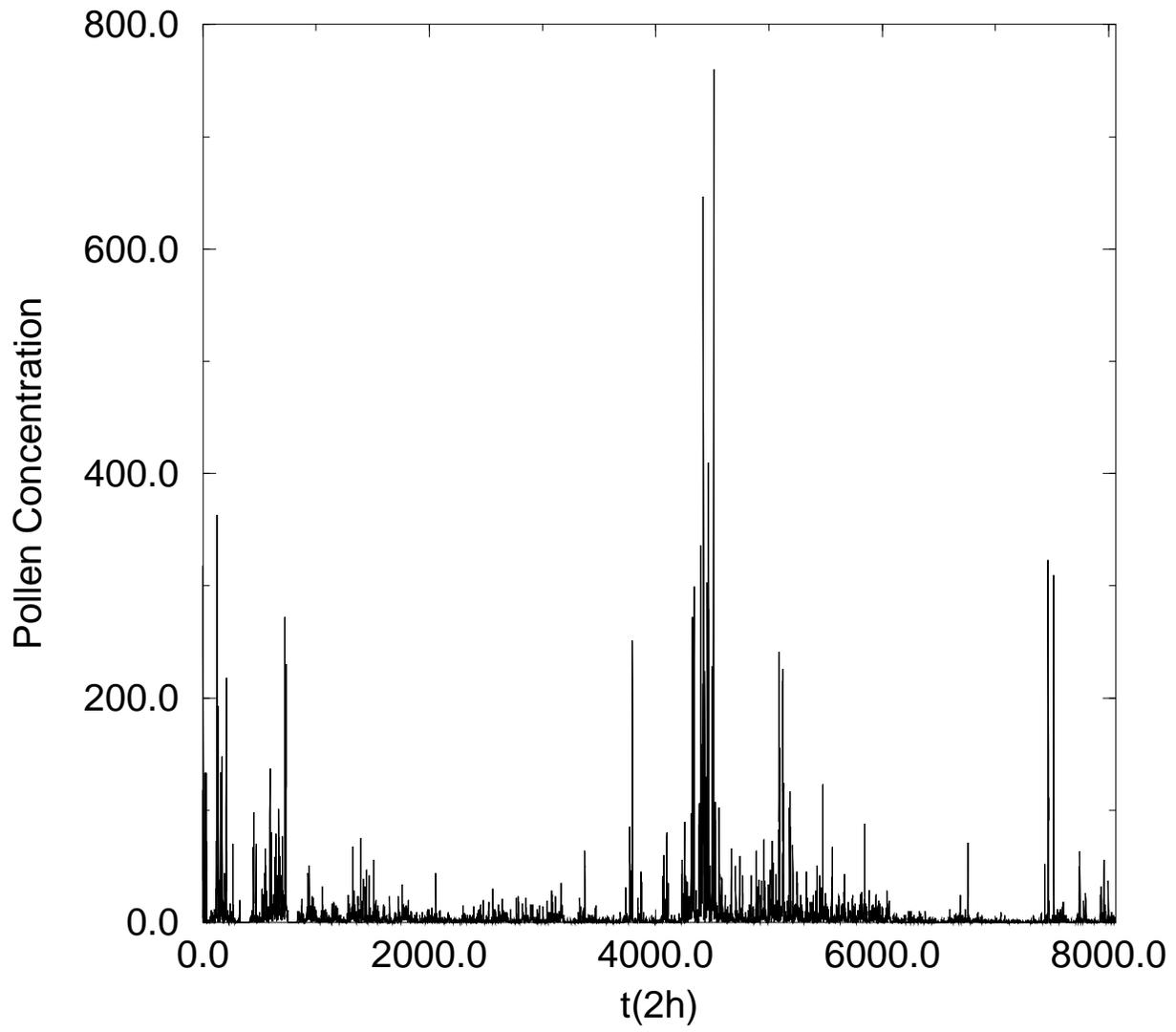
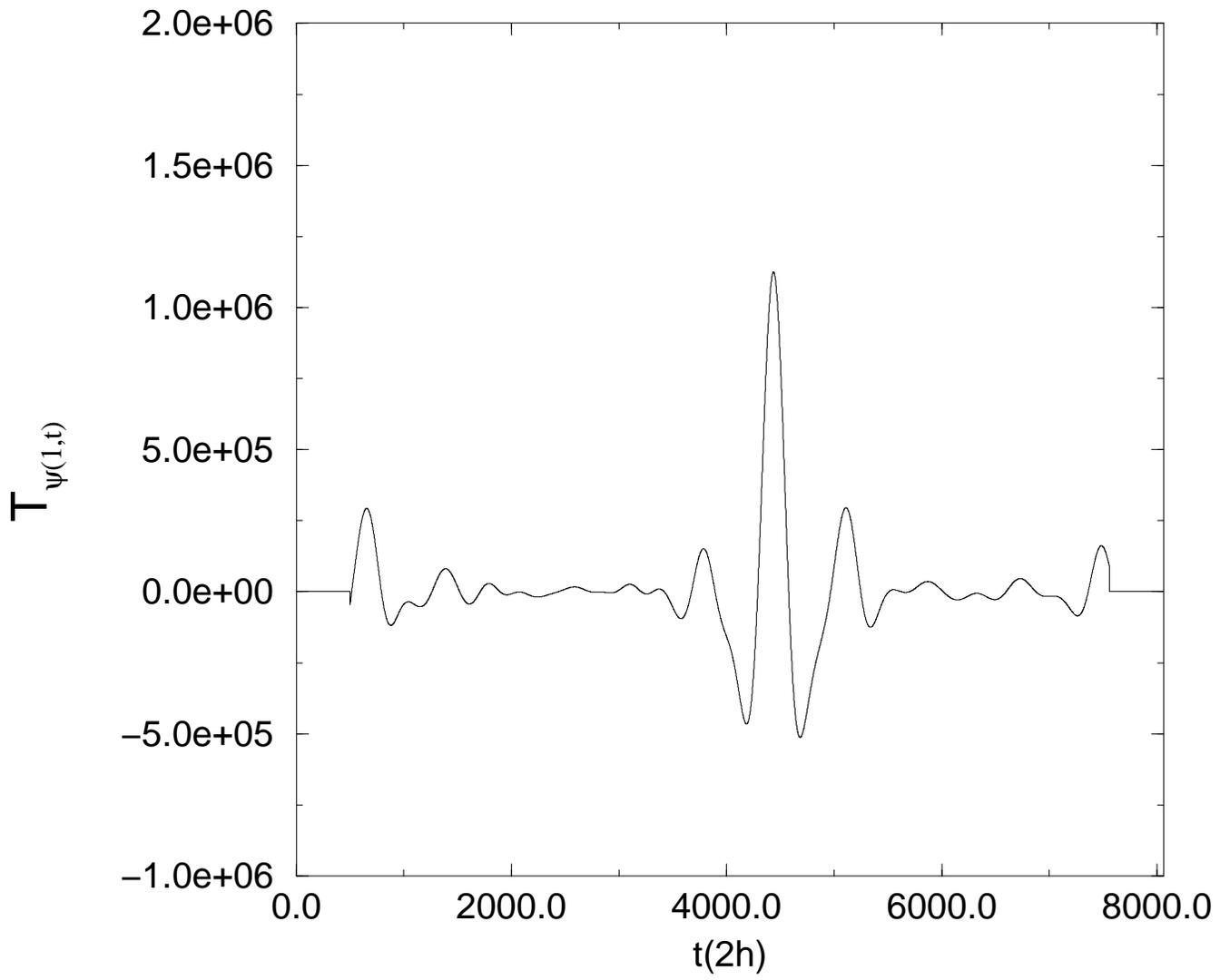
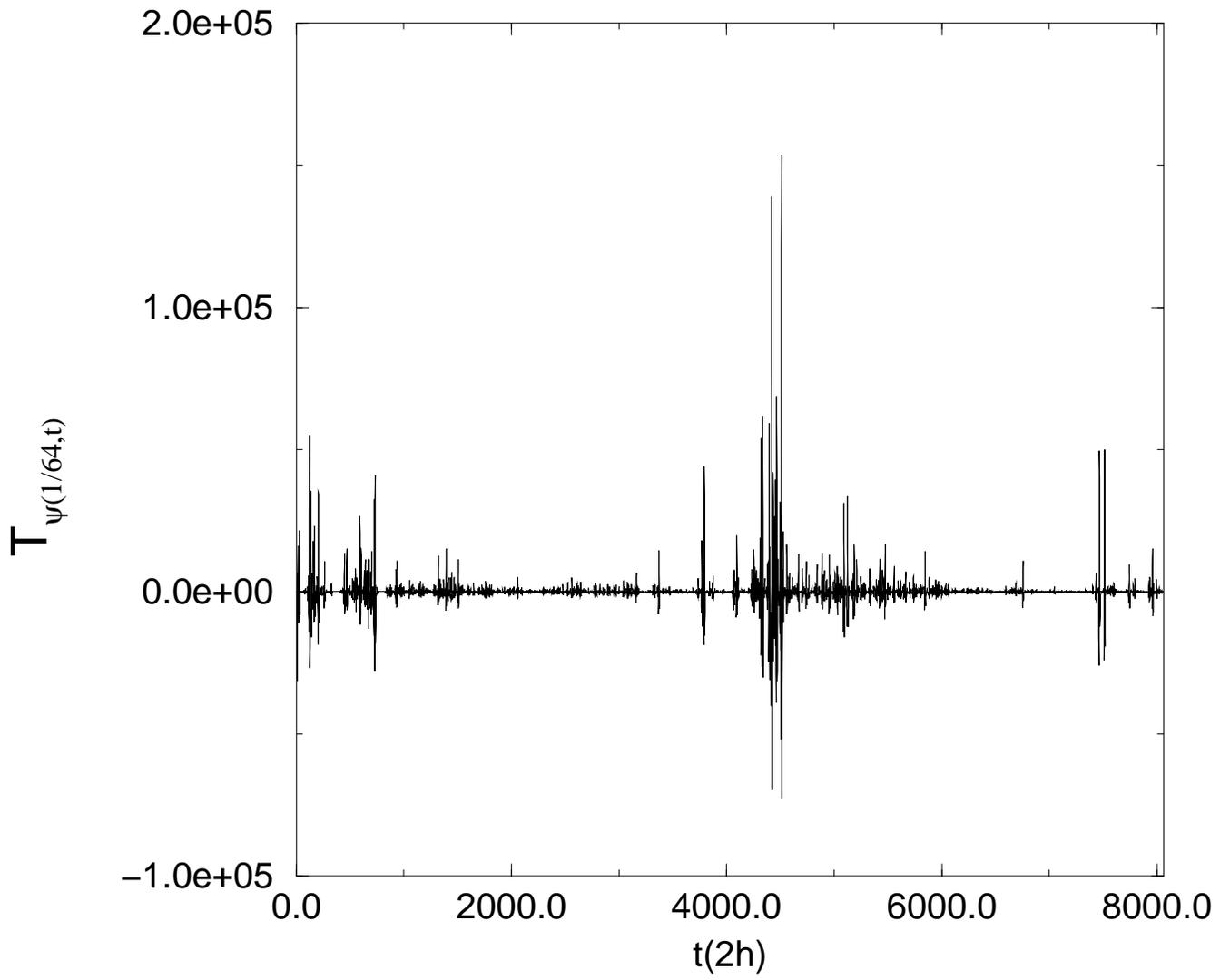
\end{document}